\documentstyle[12pt]{article}

\begin{document}
\title{\large{\bf Tricritical behaviour of the frustrated Ising antiferromagnet\\
on the honeycomb lattice}}
\author{ {\bf A. Bob\'ak}$\,^a$
\footnote{Corresponding author. Fax: +421-55-6222124.
\protect\newline
\hspace*{0.3cm} \it {E-mail address:} andrej.bobak@upjs.sk
(A. Bob\'ak).},
\hspace*{0.5mm}
{\bf T. Lu\v{c}ivjansk\'y}$\,^{a,b}$, {\bf M. \v{Z}ukovi\v{c}}$\,^a$, {\bf M. Borovsk\'y}$\,^a$, \\
{\bf T. Balcerzak}$\,^c$, 
\\
\normalsize $^a\,$Department of Theoretical Physics and Astrophysics,
Faculty of Science,\\
\normalsize P. J. \v{S}af\'{a}rik University, Park Angelinum 9, 041  54 Ko\v{s}ice, Slovak Republic\\
\normalsize $^b\,$Fakult\"at f\"ur Physik, Universit\"at Duisburg-Essen, D-47048 Duisburg, Germany\\
\normalsize $^c\,$Department of Solid State Physics, University of
\L\'{o}d\'{z}, \\
\normalsize Pomorska 149/153, 90-236 \L\'{o}d\'{z}, Poland\\
}
\date{}
\maketitle
\vspace*{1cm}

{\bf Abstract}

\hspace*{0.5cm}We use the effective-field theory with correlations based on different cluster sizes to investigate phase diagrams of the frustrated Ising antiferromagnet on the honeycomb lattice with isotropic interactions of the strength $J_1 < 0$ between nearest-neighbour pairs and $J_2 < 0$ between next-nearest neighbour pairs of spins. We present results for the ground-state energy as a function of the frustration parameter $R = J_2/|J_1|$. We find that the cluster-size has a considerable effect on the existence and location of a tricritical point in the phase diagram at which the phase transition changes from the second order to the first one.

\vspace*{0.25cm}

\newpage
{\bf 1. Introduction} \\

\hspace*{0.5cm}Since a honeycomb lattice antiferromagnet with only nearest-neighbour exchange interactions ($J_1$) is considered as a bipartite lattice, the ground state exhibits long-range ordering. The system becomes frustrated like the square lattice, if the next-nearest-neighbour exchange interactions ($J_2$) are considered. However, spin fluctuations are expected to be larger for the honeycomb lattice than the square lattice because the coordination number $z=3$ in the honeycomb lattice is smaller than that of $z=4$ in the square lattice. Hence, it is interesting to study the magnetic ordering on the honeycomb lattice under frustrating interactions. \\
\hspace*{0.5cm}We note that investigations of the frustrated two-dimensional Ising antiferromagnet (AF) with spin-$\frac{1}{2}$ on a square lattice have a long history (see, e.g. $[1-8]$). In particular, it has been found that the introduction of competing interactions is accompanied by the appearance of new ground states at the critical point $R\equiv J_2/|J_1| = -0.5$ and due to the ground-state degeneracy there is no long-range order at finite temperatures $[4, 9-11]$. Despite the simplicity of the model, it has been proved difficult to precisely determine the order of the phase transition. Now, it is well established by using different approximate studies $[10-12]$ and the Monte Carlo method $[13-18]$ that in the region of $R<-0.5$, the phase transition changes at a tricritical temperature from the second order to the first order. However, a very recent cluster mean-field calculation $[18]$ with a cluster of the size $4\times 4$ and the effective-field theory with correlations based on the different cluster sizes $[19]$ give change in the order of the phase transition not only for $R < -0.5$ but also in the region of $R > -0.5$. \\
\hspace*{0.5cm}Interestingly, a similar attention has not been paid so far to the frustrated Ising AF with spin-$\frac{1}{2}$ on the honeycomb lattice. A special feature of this lattice is that it is not a Bravais lattice, i.e., a translation invariance of the full lattice is broken for any type of state $[20]$. This non-Bravais lattice can be viewed as a composition of two interlacing triangular sublattices and the lattice is constructed by two vectors of the triangular Bravais lattice (see Fig. 1 in $[21]$). Hence, for a transition from a paramagnetic state to a magnetically ordered phase, the spatial symmetry is not reduced as for the square lattice. We expect that the non-Bravais character of this bipartite lattice results in a behaviour that cannot be observed in the square lattice or other Bravais lattices $[22]$. Moreover, in view of recent experimental activities $[23-28]$, materials regarded as various types of spin systems on honeycomb lattices are expected to be synthesized. \\
\hspace*{0.5cm}Motivated by the above considerations, in this paper we investigate the phase diagram and critical properties of the frustrated $J_1-J_2$ Ising AF on the honeycomb lattice. As far as we know, this model has not been analyzed in the literature. An interest in the honeycomb lattice is also promoted in recent years because of its relevance to graphen $[29]$. However, when second-neighbor interactions are taken into account or when a magnetic field is applied to the honeycomb lattice, the Hamiltonian is no longer exactly solvable and only approximate analytical studies or numerical approaches are possible to attack this more general problem.  \\
\hspace*{0.5cm}In this paper we employ the effective-field theory with correlations (EFT) based on different cluster sizes which has been used for an investigation of frustration in the square case $[19]$. Therefore, it will be interesting to compare effects of frustration on the phase diagram of these bipartite lattices. This approach is based on the differential operator technique introduced into exact Ising spin identities and has been successfully applied to a variety of spin-$\frac{1}{2}$ and higher spin problems (for a review see, e.g., Ref. $[30, 31]$) including a geometrically frustrated triangular lattice Ising AF $[32-34]$. Namely, here we will study the frustrated $J_1-J_2$ Ising AF on the honeycomb lattice in its parameter space using EFT based on one-, two-, four-, and six-spin clusters. It is important that the present EFT allows us to treat large clusters in a simpler and more efficient computational manner.   \\


{\bf 2. Theory} \\
\hspace*{0.5cm}We consider the frustrated honeycomb Ising AF with competing nearest-neighbour $(J_1 < 0)$ and next-nearest-neighbour $(J_2 < 0)$ interactions. The Hamiltonian of the model is given by 
\begin{equation}
\label{ham}
H = -J_1\sum_{\langle i,j\rangle}s_i s_j - J_2 \sum_{\langle i,i_{2}\rangle}s_i s_{i_2},
\end{equation}
with $s_i = \pm 1$, where the first and second sums are taken over all pairs of nearest-neighbours (nn) and next-nearest-neighbours (nnn) of spins, respectively. \\
\hspace*{0.5cm}Before calculation of the transition line between ordered and paramagnetic phases, it is appropriate to first consider the ground state of this model. For $J_2 = 0$ the ground state of the Hamiltonian (\ref{ham}) is the known AF solution with the energy per site $E_{AF}/N = -3/2 |J_1|$. However, adding the nnn AF interactions yields an increase of the ground state energy per site for the AF state:  
\begin{equation}
\label{antif}
\frac{E_{AF}}{N} = -\frac{3}{2} (|J_1| + 2 J_2).
\end{equation}
In this case each site has its three nn on the other sublattice and six nnn on its own sublattice. For a large negative $J_2$ the system orders in the collinear striped states (CS) described either by alternate single ferromagnetic columns of antiparallel spins (Fig.~$1$(a)) or alternate pairs of columns consisting of AF coupled spins (Fig.~$1$(b)) (see Ref. $[35, 36]$). In such case the ground state is degenerate and its energy per site is given by 
\begin{equation}
\label{strip}
\frac{E_{CS}}{N} = -\frac{1}{2}(|J_1| - 2 J_2).
\end{equation}
A critical point separating these ordered phases is located at $R_c = -1/4$, where the transition temperature is suppressed to $T = 0$ K. This value may be compared to that of the frustrated $J_1-J_2$ Ising model on the square lattice $R_c = -1/2$, where the energy of the collinear (or superantiferromagnetic) state depends only on the value of $J_2$ coupling $[14]$. Due to the degeneracy of the ground state the system remains disordered at all finite temperatures for $R < -1/4$. Therefore, we focus only on the AF phase which exists for $R > -1/4$.\\
\hspace*{0.5cm}A starting point of the EFT for our Ising spin system is generalized Callen-Suzuki $[37, 38]$ exact identity  
\begin{equation}
\label{identitygeneral}
{\langle O_{\{{n}\}}\rangle} = {\Bigg\langle \frac{\rm{Tr_{\{n\}}}[\it O_{\{{n}\}} \exp(-\beta {\it H_{\rm \it \{n\}}})]}{\rm{Tr_{\{n\}}}[\exp(-\beta {\it H_{\rm \it \{n\}}})]}\Bigg \rangle}, 
\end{equation}  
where the partial trace $\rm{Tr_{\{n\}}}$ is to be taken over the set $\{{n\}}$ of spin variables specified by the cluster spin Hamiltonian $\it H_{\rm \it\{n\}}$. Here, $\it O_{\{n\}}$ denotes any arbitrary spin function including the set of all $\{{n\}}$ spin variables (finite cluster) and $\langle \cdots \rangle$ denotes the usual thermal average.\\

{\it 2.1. Single-spin cluster approach}\\
\newline{}
\hspace*{0.5cm}Let us consider first the cluster containing only one spin on site $i$ and $A$ sublattice which interacts with other nn and nnn spins from the neighbourhood. In this approach the multispin Hamiltonian $\it H_{\rm \it\{n\}}$ for the AF single-spin cluster ($n = 1$) on the honeycomb lattice is given by
\begin{equation}
\label{hampartone}
H_{\{{1}\}}^{AF} = -s_i^A h_i^{AF},
\end{equation}
with
\begin{equation}
\label{one}
h_i^{AF}= J_1\sum_{i_1=1}^{3}s_{i_1}^B  + J_2 \sum_{i_2=1}^{6}s_{i_2}^A, 
\end{equation}
where $s_i^A$ and $s_j^B$ are spin variables on sublattices $A$ and $B$, respectively, and the superscript $\rm{AF}$ denotes the antiferromagnetic system. After performing the trace over the selected spin $s_i^A$ on the right-hand side of the relation (\ref{identitygeneral}), applying the differential operator technique, and using the van der Waerden identity for the two-state Ising spin system, one finds  
\begin{equation}
\label{exactmagA}
m_A \equiv \langle s_i^A \rangle = {\Bigg\langle \prod_{i_1=1}^{3}(A_1 + B_1 s_{i_1}^B) \prod_{i_2=1}^{6}(A_2 + B_2 s_{i_2}^A) \Bigg \rangle}\tanh(\beta x)\Big|_{x=0}, 
\end{equation}
where $A_\nu = \cosh(J_\nu D_x)$, $B_\nu = \sinh(J_\nu D_x)$ $(\nu = 1, 2)$, and $D_x = \partial/\partial x$ is the differential operator. \\
\hspace*{0.5cm}To proceed further, one has to approximate the thermal multiple correlation functions occurring on the right-hand side of Eq. (\ref{exactmagA}) as follows:
\begin{equation}
\label{approx}
{\langle s_{i_1}^B s_{i'_1}^B \cdots s_{i_2}^A\rangle} \approx{\langle s_{i_1}^B\rangle}{\langle s_{i'_1}^B \rangle} \cdots {\langle s_{i_2}^A \rangle},  
\end{equation}    
which means that nn and nnn of site $i$ are assumed to be completely independent of each other. It should be noted here that the approximation (\ref{approx}) is quite superior to the standard mean-field theory since even though it neglects correlations between different spins but takes the single-site kinematic relations exactly into account through the van der Waerden identity. Based on this approximation, Eq. (\ref{exactmagA}) reduces to  
\begin{equation}
\label{magone_A}
m_A = (A_1 + B_1 m_B)^3(A_2 + B_2 m_A)^6\tanh(\beta x)\Big|_{x=0}, 
\end{equation}
where $m_\alpha$ ($\alpha = A, B$) are the sublattice magnetizations per site. At this place, in order to solve the problem generally, we need to evaluate the sublattice magnetization $m_B$. It can be derived in the same way as $m_A$ by the use of (\ref{identitygeneral}) for the selected spin $s_j$ on $B$ sublattice. However, at zero magnetic field we have $m_{AF} \equiv m_A = -m_B$ and the equation for $m_B$ is the same as Eq. (\ref{magone_A}). Therefore, in what follows we use only Eq. (\ref{magone_A}), which in this case takes the final form 
\begin{equation}
\label{magone_{AF}}
m_{AF} = \sum_{{n}=0}^{4} K_{2n+1}^{AF} m_{AF}^{2n+1}, 
\end{equation}
where the coefficients $K_{2n+1}^{AF}$, which depend on $T$ and $R$, can be easily calculated within the symbolic programming by using the mathematical relation $\exp( \lambda D_x) f(x) = f(x+\lambda)$. Because the final expressions for these coefficients are lengthy, their explicit form is omitted.  \\
\hspace*{0.5cm}We are now interested in studying the transition temperature (or the phase diagram) and the tricritical point of the model where the transition changes from the second order to the first order. In the neigbourhood of a second-order transition line where the order parameter $m_{AF}$ is small, Eq. (\ref{magone_{AF}}) can be rewritten as 
\begin{equation}
\label{transition_one}
m_{AF} = K_1^{AF} m_{AF} +  K_3^{AF} m_{AF}^3 + \cdots. 
\end{equation}
The second-order phase transition line is then determined by the conditions  
\begin{equation}
\label{transit_single}
K_1^{AF} = 1 \quad \quad {\mathrm and} \quad \quad K_3^{AF} < 0. 
\end{equation} 
Note that we have verified that the coefficient $K_3^{AF}$ is negative in the entire $(T, R)$ plane for $R > -1/4$. Thus, within the present EFT based on the single-spin cluster we have only a second-order transition line between the AF and paramagnetic (P) phases. \\     
\newpage{}
{\it 2.2. Multi-spin cluster approach}\\
\newline{}
\hspace*{0.5cm}In order to take into account effects of frustration within the present EFT more precisely, it is necessary to consider at least a two-spin cluster. In this approach, we select two nn spins, labeled $i$ and $j$, which interact with other nn and nnn spins from the neighborhood $[39]$. Hence, the multi-spin Hamiltonian $\it H_{\rm \it\{n\}}$ for the AF two-spin cluster ($n=2$) on the honeycomb lattice (Fig.~$2$) is given by 
\begin{equation}
\label{hamparttwo}
{\it H_{\rm \it \{ij\}}}^{AF} = -J_1 s_i^As_j^B - s_i^A h_i^{AF} - s_j^B h_j^{AF}, 
\end{equation}
with 
\begin{equation}
\label{two}
h_i^{AF} = J_1\sum_{i_1=1}^{2}s_{i_1}^B + J_2 \sum_{i_2=1}^{6}s_{i_2}^A, \quad \quad  h_j^{AF} = J_1\sum_{j_1=1}^{2}s_{j_1}^A + J_2 \sum_{j_2=1}^{6}s_{j_2}^B,
\end{equation}  
where the terms $i_1=j$ and $j_1=i$ are excluded from summations over the indices $i_1$ and $j_1$, respectively. At this point one should notice that the neighbourhood of the sites $i$ and $j$ of the two-spin cluster for the $J_1-J_2$ model on a honeycomb lattice contains a set of common spins, namely the spins at the sites labeled by $(i_1, j_2)$ or $(j_1, i_2)$ in Fig.~$2$. These spins interact with spins of the cluster and are frustrated directly within the two-spin cluster theory, which is not the case of the one-spin cluster approximation. Now, taking this into account and using the same procedure as for the single-spin cluster, one derives the equation analogous to Eq. (\ref{magone_A}), which now reads
\begin{eqnarray}
\label{magtwo}
m_{AF} &=& \Big[A_x(1) A_y(2) + B_x(1) B_y(2) + m_B \Big(A_x(1)B_y(2)+A_y(2)B_x(1)\Big)\Big]^2 \nonumber \\
& & \times \Big[A_y(1) A_x(2) + B_y(1) B_x(2) + m_A \Big(A_y(1)B_x(2)+A_x(2)B_y(1)\Big)\Big]^2 \nonumber \\
& & \times \Big[\Big(A_x(2) + m_A B_x(2)\Big)\Big(A_y(2) + m_B B_y(2)\Big)\Big]^4f_{AF} (x,y)\Big|_{x=0,y=0}, 
\end{eqnarray}
where $A_\mu(\nu) = \cosh(J_\nu D_\mu)$, $B_\mu(\nu) = \sinh(J_\nu D_\mu)$ ($\nu = 1, 2$), $D_\mu = \partial/\partial \mu$ ($\mu = x, y$) are the differential operators and function $f_{AF}(x, y)$ is defined by 
\begin{equation}
\label{functwo}
f_{AF} (x, y) = \frac{\sinh\beta(x-y)}{\cosh\beta(x-y) + e^{2\beta J_1}\cosh\beta(x+y)}. 
\end{equation} 
Now, by using the condition $m_{AF} \equiv \langle (s_i^A-s_j^B)/2\rangle = m_A = -m_B$, Eq. (\ref{magtwo}) can be finally recast in the form 
\begin{equation}
\label{magtwo_{AF}}
m_{AF} = \sum_{{n}=0}^{5} L_{2n+1}^{AF} m_{AF}^{2n+1}, 
\end{equation}
where the coefficients $L_{2n+1}^{AF}$, which depend on $T$ and $R$, can be again easily calculated within the symbolic programming by using the mathematical relation $\exp (\lambda D_x + \gamma D_y) f_{AF}(x, y) = f_{AF}(x+\lambda, y+\gamma)$. We also note that in obtaining Eq.~(\ref{magtwo}) we have made use of the fact that $f_{AF}(x, y) = - f_{AF}(-x, -y)$ and therefore only odd differential operator functions give nonzero contributions. \\
\hspace*{0.5cm}The second-order phase transition line is then determined by  
\begin{equation}
\label{transit_two}
L_1^{AF} = 1 \quad \quad {\mathrm and} \quad \quad L_3^{AF} < 0. 
\end{equation}  
In the vicinity of the second-order phase transition line, the order parameter $m_{AF}$ is given by 
\begin{equation}
\label{trans_magn_two}
m_{AF}^2 = \frac{1-L_1^{AF}}{L_3^{AF}}. 
\end{equation}
The right-hand side of Eq. (\ref{trans_magn_two}) must be positive. If this not the case, the transition is of the first order, and hence the point at which    
\begin{equation}
\label{tricrit_two}
L_1^{AF} = 1 \quad \quad {\mathrm and} \quad \quad L_3^{AF} = 0
\end{equation}  
is the tricritical point (TCP) $[40]$. \\
\hspace*{0.5cm}To get a more convincing evidence for the existence a TCP in the phase diagram, we have also considered four- and six-spin clusters. However, analytical calculations for such large clusters would have been very lengthy and tedious, therefore, the results were obtained in a completely numerical way within the symbolic programming by using Mathematica software package $[41]$. It should be noted here that the calculation times for the large clusters become rather long even using the symbolic programming. Therefore, the highest approximation used to study the frustrated Ising honeycomb lattice is the one based on the six-spin cluster. \\
 
{\bf 3. Results and discussion} \\
\hspace*{0.5cm}Numerical results for the critical temperature $k_BT_N/|J_1|$ versus $R$ for various cluster sizes are shown in Fig.~$3$. In this figure the solid lines indicate the second-order phase transitions and the black circles denote the positions of TCPs at which the phase transitions change from the second to the first order. \\
\hspace*{0.5cm}First, by solving Eq.~(\ref{transit_single}) numerically, we obtain a phase diagram between the AF and P phases in the $(R, T)$ plane for the single-spin cluster. In this case, as seen from Fig.~$3$, the corresponding AF-P transition ($n =1$) is always of the second order and the critical temperature gradually reduces from the value $k_BT_N/|J_1| =2.1038$ at $R = 0$ to $k_BT_N/|J_1| =0$ at $R_c =-1/4$, as expected from the ground-state arguments.  \\
\hspace*{0.5cm}However, when a larger cluster than the single-spin one is used, the second-order transition line for the frustrated $J_1-J_2$ Ising honeycomb lattice terminates at the TCP. Thus, for the larger clusters there are second-order as well as first-order transitions. An example of such a phase diagram, obtained by solving Eqs.~(\ref{transit_two}) and (\ref{tricrit_two}) numerically for the two-spin cluster, is shown in Fig.~$3$ ($n = 2$). In this figure we show also a phase diagram for the four-spin cluster ($n = 4$) obtained within the present EFT in a completely numerical way. The scheme of this cluster, which consists of the spins $s_i^A, s_j^B, s_k^B$, and $s_l^B$, is illustrated in Fig.~$4$. It is seen from the figure that, similar to the two-spin cluster approximation, the corresponding 'fields' $h_i^{AF}$, $h_j^{AF}$, $h_k^{AF} $, and $h_l^{AF}$ of the four-spin cluster contain a set of common spins, namely the spins at the sites labeled by two indices in Fig.~$4$. (These 'fields', for brevity, are not presented explicitly here.) Further, it is seen from Fig.~$4$ that within the four-spin cluster approximation we take into account exactly three nn interactions $J_1$ and three nnn interactions $J_2$. For the nonfrustrated model ($R = 0$), we find that values of $k_BT_N/|J_1|$ are $1.9869$ and $1.9261$ for the two- and four-spin clusters, respectively, which indicates a relatively slow convergence to the exact value of $k_BT_N/|J_1| = 1.5186$ with the increasing cluster size. On the other hand, our estimates for the coordinates of the TCP ($k_BT_t/|J_1|; R_t$) are $(1.1352; -0.0907)$ and $(1.0820; -0.9125)$ for the two- and four-spin cluster approximations, respectively. Thus, the cluster-size has a considerable effect on the existence and location of the TCP at which the phase transition between the AF and P phases changes from the second order to the first one. We note here that the first-order transition line is not possible to calculate on the basis of Eq.~(\ref{transit_two}) since then we are not allowed to linearize Eq.~(\ref{magtwo_{AF}}) in the vicinity of the transition point. To solve this problem, one needs to calculate the free energy for the AF and P phases and to find a point of intersection. Since only an approximate expression exists for the free energy at finite temperature in the frame of the EFT based on any spin cluster (see, e.g. $[12, 19]$), we have confined our calculations only to the second-order phase transitions, including the TCP.  \\
\hspace*{0.5cm}To further investigate this tricritical behaviour, we determine the phase diagram for the six-spin cluster. We note that the choice of a six-spin cluster is not unambiguous. Indeed, one can choose a cluster with six spins in the form of a 'dumbbell' or hexagon (see Fig.~$5$). In this case the neighbourhood of the sites $i, j, k, l, m$, and $n$ of the six-spin cluster contains a set of common spins for both the 'dumbbell' spin- and the hexagon spin-clusters. In Fig.~$5$ these spins are labeled at the sites by two or three indices. It is worth noticing that in the hexagon-spin cluster there are only sites labeled by three indices, contrary to the 'dumbbell'-spin cluster where sites with two or three indices exist. Since the phase diagrams for these clusters are qualitatively the same, in Fig.~3 we show only the corresponding phase diagram for the hexagon-spin cluster. In particular, we have found that the critical temperature for $R =0$ is $k_BT_N/|J_1|=1.9064$ ('dumbbell'-spin cluster) and $k_BT_N/|J_1|=1.8673$ (hexagon-spin cluster). By comparing these values of $k_BT_N/|J_1|$ to the exact value ($k_BT_N/|J_1| = 1.5186$), it can be seen that the EFT based on the spin cluster in the form of the hexagon produces a larger improvement in the $k_BT_N/|J_1|$ than that for the 'dumbbell' cluster. This is not surprising because the present treatment based on the hexagon cluster approximation takes into account exactly six nn interactions while the EFT based on the 'dumbbell' cluster takes into account exactly only five nn interactions between the pair of spins defining the cluster (see Fig.~$5$). Therefore, a further improvement to the theory is possible if clusters with a larger number of nn interactions are considered. This is, as mentioned above, a difficult task due to the fact that calculation times for large clusters become rather long even using the symbolic programming. Finally, we estimate coordinates of the TCP at the $(R, T)$ plane, which are $(1.0528; -0.0969)$ and $(0.9189; -0.1133)$ for the 'dumbbell'- and hexagon-spin cluster approximations, respectively. Generally it is seen that within the present approach the TCP occurs at a fractionally higher negative value of the frustration parameter with an increasing cluster size, but at temperature considerably lower.  \\

{\bf 4. Conclusions} \\
\hspace*{0.5cm}We have studied the phase diagram in the (R, T) plane of the frustrated $J_1-J_2$ Ising model with spin-$\frac{1}{2}$ on a honeycomb lattice using the EFT based on different cluster sizes. We have determined that the ground-state is the AF phase for $R >-1/4$, while the system orders in the CS phase for $R < -1/4$. However, for $R < -1/4$, we have not found a long-range order at $T \neq 0$ K due to the degeneracy of the ground state. This behaviour has been also confirmed by our preliminary Monte Carlo calculations. \\
\hspace*{0.5cm}Further, in the AF region ($R > -1/4$), we have found the phase transition line between the AF and P phases. However, the present EFT predicts the TCP in the phase diagram only for clusters $n > 1$, but not for the single-spin $(n=1)$ cluster where only the second-order phase transition was observed. Since by using larger and larger clusters, better results are expected, we are forced to conclude that the frustrated $J_1-J_2$ Ising system on a honeycomb lattice exhibits the TCP in the phase diagram between the AF and P phases. Therefore, we believe that our effective-field results are qualitatively correct and the tricritical behaviour is due to stronger effects of frustration for the clusters $n > 1$ than for the single-spin $(n=1)$ cluster. A thorough Monte Carlo study or more reliable calculations for this frustrated $J_1-J_2$ model would be desirable. To our knowledge, no such studies have been attempted yet. \\

{\bf Acknowledgment} \\
\hspace*{0.5cm}This work was supported by the Scientific Grant Agency of Ministry of Education of Slovak Republic (Grant VEGA No. 1/0331/15).\\

\newpage{}

\newpage
{\bf Figure captions} \\

{\bf Figure 1:} Ground-state configurations of the $J_1-J_2$ Ising model on the honeycomb lattice showing two, (a) and (b), degenerate collinear striped states. Two sublattices are marked by black and white circles.      \\
{\bf Figure 2:} Ground-state configurations of the $J_1-J_2$ Ising model on the honeycomb lattice showing aniferromagnetic states for the two-spin cluster approximation defined by the Hamiltonian (\ref{hamparttwo}) with spins $s_i$ and $s_j$ (thick line). The sites occupied by spins that interact with one or two spins of the cluster are labeled by one or two corresponding indices, respectively. Two sublattices are marked by black and white circles.\\ 
{\bf Figure 3:} Phase diagram in the coupling-temperature plane for the $J_1-J_2$ Ising model on the honeycomb lattice based on the one- ($n=1$), two- ($n=2$), four- ($n=4$), and six-spin ($n=6$) clusters. The latter cluster corresponds to the hexagon-spin one (see also text). The solid lines indicate second-order transitions and the black circles denote the position of a tricritical point. AF and $P$ are the antiferromagnetic and paramagnetic phases. \\
{\bf Figure 4:} Ground-state configurations of the $J_1-J_2$ Ising model on the honeycomb lattice showing aniferromagnetic states for the four-spin cluster approximation with spins $s_i, s_j, s_k$, and $s_l$ (thick lines). The sites occupied by spins which interact with one or two spins of the cluster are labeled by one or two corresponding indices, respectively. Two sublattices are marked by black and white circles. \\
{\bf Figure 5:} Two options of the six-spin cluster with spins $s_i, s_j, s_k, s_l, s_m$, and $s_n $ for the antiferromagnetic arrangement on the honeycomb lattice (thick lines): (a) for the 'dumbbell'-spin cluster and (b) for the hexagon-spin cluster. The sites occupied by spins which interact with one, two or three spins of the cluster are labeled by one, two or three corresponding indices, respectively. Two sublattices are marked by black and white circles. \\


\begin{thebibliography}{12}
\bibitem{}
L. P. Kadanoff, Phys. Rev. Lett. 39 (1977) 903.
\bibitem{}
M. P. Nightingale, Phys. Lett. 59A 468 (1977) 468.
\bibitem{}
R.H. Swendsen, S. Krinsky, Phys. Rev. Lett. 43 (1979) 177.
\bibitem{}
P. Landau, Phys. Rev. B 21 (1980) 1285.
\bibitem{}
K. Binder, D.P. Landau, Phys. Rev. B 21 (1980) 1941.
\bibitem{}
D.P. Landau, K. Binder, Phys. Rev. B 31 (1985) 5946.
\bibitem{}
J. Oitmaa, J. Phys. A: Math. Gen. 14 (1981) 1159.
\bibitem{}
H.W.J. Bl\"{o}te, A. Compagner, A. Hoogland, Physica 141A (1987) 375.
\bibitem{}
M.D. Grynberg, B. Tanatart, Phys. Rev. B 45 (1992) 2876.
\bibitem{}
J.L. Mor\'an-L\'opez, F. Aguilera-Granja, J.M. Sanchez, Phys. Rev. B 48 (1993) 3519. 
\bibitem{}
J.L. Mor\'an-L\'opez, F. Aguilera-Granja, J.M. Sanchez, J. Phys.: Condens. Matter 6 (1994) 9759. 
\bibitem{}
R.A. dos Anjos, J.R. Viana, J.R. de Sousa, Phys. Lett. A 372 (2008) 1180.
\bibitem{}
E. L\'opez-Sandoval, J.L. Mor\'an-L\'opez, F. Aguilera-Granja, Solid State Commun. 112 (1999) 437.
\bibitem{}
A. Kalz, A. Honecker, S. Fuchs, and T. Pruschke, Eur. Phys. J. B 65 (2008) 533.
\bibitem{}
A. Kalz, A. Honecker, S. Fuchs, T. Pruschke, J. Phys.: Conf. Ser. 145 (2009) 012051.
\bibitem{}
A. Kalz, A. Honecker, M. Moliner, Phys. Rev. B 84 (2011) 174407.
\bibitem{}
S. Jin, A. Sen, A.W. Sandvik, Phys. Rev. Lett. 108 (2012) 045702.
\bibitem{}
S. Jin, A. Sen, W. Guo, A.W. Sandvik, Phys. Rev. B 87 (2013) 144406.
\bibitem{}
A. Bob\'ak, T. Lu\v{c}ivjansk\'y  Borovsk\'y, M. \v{Z}ukovi\v{c}, Phys. Rev. E 91 (2015) 032145.
\bibitem{}
A. Mattsson, P. Fr\"{o}jdh, T. Einarsson,  Phys. Rev. B 49 (1994) 3997.
\bibitem{}
H. Mosadeq, F. Shahbazi, S.A. Jafari, J. Phys.: Condens. Matter 23 (2011) 226006.
\bibitem{}
I. Affleck Phys. Rev. B 37 (1988) 5186.
\bibitem{}
V. Kataev, A. M\"{o}ller, U. L\"{o}w, W. Jung, N. Schittner, M. Kriener, A. Freimuth, J. Magn. Magn. Mater. 290-291 (2005) 310. 
\bibitem{}
Y. Miura, R. Hiari, Y. Kobayashi, M. Sato, J. Phys. Soc. Japan 75 (2006) 084707.
\bibitem{} 
O. Smirnova, M. Azuma, N. Kumada, Y. Kusano, M. Matsuda, Y. Shimakawa, T. Takei, Y. Yonesaki, N. Kinomura, J. Am. Chem. Soc. 131 (2009) 8313.
\bibitem{} 
S. Okubo, F. Elmasry, W. Zhang, M. Fujisawa, T. Sakurai, H. Ohta, M. Azuma, O.A. Sumirnova, N. Kumada, J. Phys.: Conf. Ser. 200 (2010) 022042.
\bibitem{}
A.A. Tsirlin, O. Janson, H. Rosner, Phys. Rev. B 82 (2010) 144416. 
\bibitem{}
M. Matsuda, M. Azuma, M. Tokunaga, Y. Shimakawa, N. Kumada, Phys. Rev. Lett. 105 (2010) 187201.
\bibitem{}
A.H. Castro Neto, F. Guinea, N.M.R. Peres, K.S. Novoselov, A.K. Geim, Rev. Mod. Phys. 81 (2009) 109.
\bibitem{}
T. Kaneyoshi, Acta Phys. Polonica A 83 (1993) 703.
\bibitem{}
J. Stre\v{c}ka, M. Ja\v{s}\v{c}ur, Acta Phys. Slovaca 65 (2015) 235.
\bibitem{}
M. \v{Z}ukovi\v{c}, M Borovsk\'y, A. Bob\'ak, Phys. Lett. A 374 (2010) 4260. 
\bibitem{}
M. \v{Z}ukovi\v{c}, M Borovsk\'y, A. Bob\'ak, J. Magn. Magn. Mater. 324 (2012) 2687.
\bibitem{}
M Borovsk\'y, M. \v{Z}ukovi\v{c}, A. Bob\'ak, Physica A 392 (2013) 157.
\bibitem{}
T. Kudo, S. Katsura, Prog. Theor. Phys. 56 (1976) 435.
\bibitem{}
S. Katsura, T. Ide, T. Morita, J. Stat. Phys. 42 (1986) 381.
\bibitem{}
H.B. Callen, Phys. Lett. 4 (1963) 161.
\bibitem{}
M. Suzuki, Phys. Lett. 19 (1965) 267.
\bibitem{}
A. Bob\'ak, M. Ja\v{s}\v{c}ur, Phys. Stat. Sol. B 135 (1986) K9.
\bibitem{}
N. Benayard, A. Benyoussef, N. Boccara, J. Phys. C 18 (1985) 1899.
\bibitem{}
Wolfram Research, MATHEMATICA, Version 9.0 (Champaign, Illinois, 2012).
\end{thebibliography}
\end{document}